%% file: 0-main.tex
\documentclass[conference]{IEEEtran}
\IEEEoverridecommandlockouts
\ifCLASSOPTIONcompsoc
  \usepackage[nocompress]{cite}
\else
  \usepackage{cite}
\fi

\ifCLASSOPTIONcompsoc
 \usepackage[caption=false,font=footnotesize,labelfont=sf,textfont=sf]{subfig}
\else
 \usepackage[caption=false,font=footnotesize]{subfig}
\fi

\usepackage{graphicx}
\usepackage{amsfonts}
\usepackage{amsmath}
\usepackage{breqn}
\usepackage{etoolbox}

\usepackage[utf8]{inputenc}
\usepackage[T1]{fontenc}
\usepackage{graphicx}
\usepackage{cite}
\usepackage{amsmath,amssymb,amsfonts}
\usepackage{graphicx}
\usepackage{textcomp}
\usepackage{xcolor}
\usepackage{float}

\usepackage{xcolor,soul}
\usepackage{xspace}
\usepackage{enumitem}
\usepackage{titlecaps}
\usepackage{tabularx}
\usepackage{balance}
\usepackage{color}
\usepackage{comment}
\usepackage{csquotes}
\usepackage{amsmath,amssymb}
\usepackage{url}
\usepackage{hyperref}

\usepackage{algorithm}% http://ctan.org/pkg/algorithm
\PassOptionsToPackage{noend}{algpseudocode}% comment out if want end's to show
\usepackage{algpseudocode}% http://ctan.org/pkg/algorithmicx
\errorcontextlines\maxdimen

\newcommand{\etal}{\textit{et al}.}
    
\begin{document}

\makeatletter
%%%%%%%%%%%%%%%%%%%%%%%%%%%%%% User specified LaTeX commands.
\def\ps@IEEEtitlepagestyle{%
  \def\@oddfoot{\mycopyrightnotice}%
  \def\@evenfoot{}%
}
\def\mycopyrightnotice{%
  {\footnotesize \textcolor{red}{\begin{tabular}[t]{@{}l@{}} This paper has been accepted for publication by the 2023 IEEE International Conference on Communications (ICC 2023). © 2023 IEEE. Personal use \\ of this material is permitted. Permission from IEEE must be obtained for all other uses, in any current or future media, including reprinting/republishing \\ this material for advertising or promotional purposes, creating new collective works, for resale or redistribution to servers or lists, or reuse of any copyrighted \\ component of this work in other works.\end{tabular}}}% <--- Change here
  \gdef\mycopyrightnotice{}% just in case
}

\title{An NDN-Enabled Fog Radio Access Network Architecture With Distributed In-Network Caching}

% An NDN-Enabled Fog Radio Access Network With Distributed In-Network Caching

\author{
\IEEEauthorblockN{Sifat Ut Taki}
\IEEEauthorblockA{University of Notre Dame \\
staki@nd.edu}
\and
\IEEEauthorblockN{Spyridon Mastorakis}
\IEEEauthorblockA{University of Notre Dame \\
mastorakis@nd.edu}
}

% \and

\makeatletter
\patchcmd{\@maketitle}
  {\addvspace{0.5\baselineskip}\egroup}
  {\addvspace{-1.5\baselineskip}\egroup}
  {}
  {}
\makeatother

\maketitle

\input{1-abstract}

\input{2-introduction}

\input{3-ndn_fran}

\input{4-sys_model}

\input{5-results}

\input{6-conclusion}

\appendices
\input{apndx1}

\section*{Acknowledgements}

This work is partially supported by the National Science Foundation through awards CNS-2104700, CNS-2016714, and CBET-2124918, as well as the National Institutes of Health through award NIGMS/P20GM109090.

\bibliographystyle{unsrt}
\bibliography{ref}

\end{document}

%% file: 1-abstract.tex
\begin{abstract}
% The need for a better Internet architecture in cellular networks is becoming more prominent as the number of smart devices are rapidly increasing every year. 

To meet the increasing demands of next-generation cellular networks (e.g., 6G), advanced networking technologies must be incorporated. On one hand, the Fog Radio Access Network (F-RAN), has been proposed as an enhancement to the Cloud Radio Access Network (C-RAN). On the other hand, efficient network architectures, such as Named Data Networking (NDN), have been recognized as prominent Future Internet candidates. Nevertheless, the interplay between F-RAN and NDN warrants further investigation. In this paper, we propose an NDN-enabled F-RAN architecture featuring a strategy for distributed in-network caching. Through a simulation study, we demonstrate the superiority of the proposed in-network caching strategy in comparison with baseline caching strategies in terms of network resource utilization, cache hits, and fronthaul channel usage.

%the possibility of an NDN-enabled F-RAN architecture with a more effective in-network cache distribution strategy is explored. To incorporate NDN with F-RAN, some design changes to the protocol stacks of the existing F-RAN components are presented at first. Next, the problem with the default NDN caching policy is demonstrated with a simple case study. Subsequently, an objective function for the proposed in-network cache distribution strategy is presented to achieve optimal efficiency in cache resource usage throughout the network, along with a proof of its NP-hardness. Moreover, an algorithm based on the cache distribution strategy for the F-RAN nodes is proposed. Through extensive simulation results, we demonstrate the superior performance gain of the proposed cache distribution strategy compared with default caching strategies in terms of better resource utilization, more cache hits, and lower fronthaul channel usage.
\end{abstract}

\begin{IEEEkeywords}
Fog radio access network, Named Data Networking, 6G, in-network caching.
\end{IEEEkeywords}

%% file: 2-introduction.tex
\ifCLASSOPTIONcompsoc
\IEEEraisesectionheading{\section{Introduction}\label{sec:intro}}
\else
\section{Introduction}
\label{sec:intro}
\fi

\subsection{Background}

%\IEEEPARstart{T}{he} redundant and inefficient transmission is a major problem on today's Internet, especially when the number of smart devices is increasing on a daily basis. According to Cisco's estimation, monthly global mobile data traffic will be 77 exabyte in 2022, which will occupy 20\% of the total IP traffic~\cite{forecast2019cisco}. 

\IEEEPARstart{T}{he} number of mobile devices that connect to the Internet grows year by year. To prepare for upcoming surges in Internet usage, it is crucial to realize robust Internet architectures for the next generations of cellular networks, such as 6G \cite{shahraki2021comprehensive}. To address this challenge, architectures, such as the Cloud Radio Access Network (C-RAN), have been proposed~\cite{7064897}. Such architectures combine cloud computing with Radio Access Networks (RANs). %However, C-RAN has some limitations and challenges (e.g., heavy fronthaul channel usage). More optimized and efficient algorithms are still being developed for heterogeneous C-RAN architectures.
C-RAN's fundamental idea is to transfer the functionality of radio resources management and signal processing from a base station to a Base Band Unit (BBU) pool on the cloud. The task of the BBU pool in C-RAN is to provide spectral and energy efficiency by utilizing cloud computing~\cite{peng2016}. C-RAN aims to decouple the functionality of base stations into a centralized BBU pool and several remote radio heads, so that more remote radio heads can be deployed in areas with high demand. %The C-RAN architecture is mostly praised for its inherent capabilities of fulfilling recent needs; however, it has its fair share of limitations and challenges. 
Nevertheless, a major limitation of the C-RAN architecture is the delay in the fronthaul link: the communication channel between the BBU pool and the remote radio heads. This delay can make the efficiency gained from collaborative processing and cooperative radio resource allocation insignificant~\cite{9210131}. %As a result, a more complete solution was needed in order to make the architecture feasible in most scenarios.

% The prospects of the upcoming 6G cellular network includes not only faster data rates (more than 10 Gbps), but also more efficient, reliable, and precise connections \cite{piran2019learning}. 6G is expected to have a connection density of 10 million/km$^2$ while achieving 3 times more SE, and 10 times more EE compared to 5G~\cite{chen2020}. Technologies like software-defined networking (SDN), network function virtualization (NFV), cloud computing (CC), fog computing (FC), and artificial intelligence (AI) are expected to be utilized more in 6G. As such, a robust radio access network architecture is necessary to meet these expected requirements.

%Fog computing  is an extension of CC. It was designed to mitigate some of the limitations of CC by moving the computation layer closer to the edge nodes in order to reduce latency and increase efficiency, which inherently reduced the dependence on the central cloud source. The application of the FC principles in networking gave birth to fog network, which allowed network edge devices to communicate with each other and process data within the edge network \cite{pham2020survey}. 

\begin{figure}[!t]
\centering
\includegraphics[width=2in]{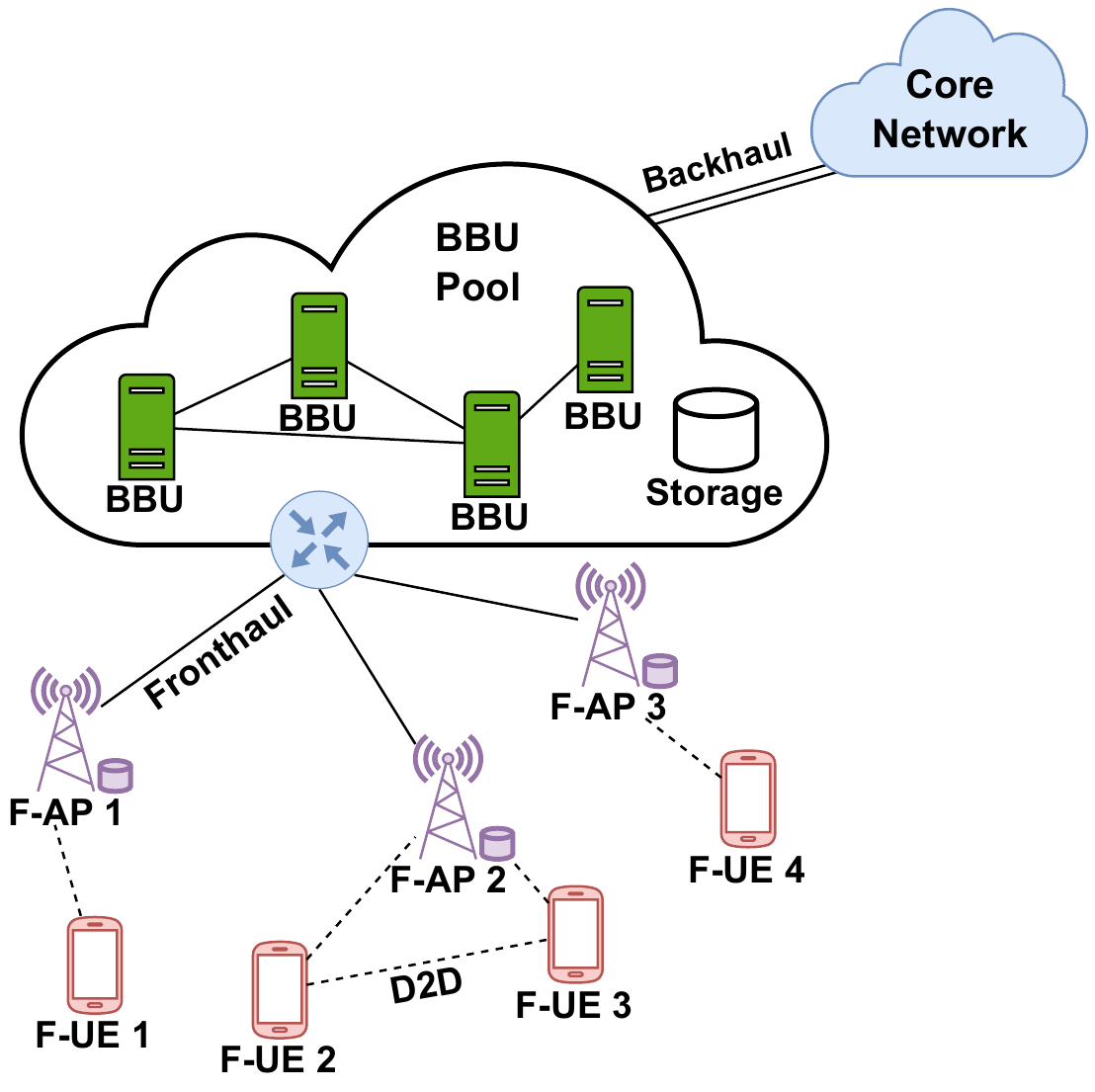}
\caption{A fog radio access network model.}
\vspace{-0.5cm}
\label{fig:fran_model}
\vspace{-0.1cm}
\end{figure}

As a solution to this limitation, Fog Radio Access Network (F-RAN) was proposed as an evolution of C-RAN, which features an integration of fog computation with radio access networks~\cite{zhang2017fran}. %When fog network was integrated into the radio access network to create F-RAN, some new terminologies were introduced with additional features. BS and user equipment (UE) were replaced with 
In the context of an F-RAN, a Fog Access Point (F-AP) and Fog User Equipment (F-UE) provide functionality equivalent to the functionality of a base station and user equipment with additional features, such as edge caching and Artificial Intelligence~\cite{FRANBookCh2}, as we illustrate in Figure~\ref{fig:fran_model}. %For example, F-APs and the BBU pool can be defined as distributed units and a central unit in 5G, respectively. 
%F-RAN reduces the burden on the fronthaul channel by allowing edge nodes in the network to store cache. 
F-RAN incorporates a number of cache-aided fog nodes in order to bring contents physically closer to users by offloading them to both F-APs and F-UEs. The main motivation behind the F-RAN architecture is to enable the BBU pool to serve more users by offloading popular contents to its F-APs and F-UEs. As a result, during the peak hours, the F-UEs can retrieve requested data from nearby F-APs or other F-UEs via Device-to-Device (D2D) communication when possible. %F-RAN is a promising candidate for 6G cellular network.

% The Internet protocol (IP)-based Internet, which relies on end-to-end communication, is currently facing some major challenges. The IP architecture is unable to adapt the modern application requirements; thus, a new concept of information-centric networking (ICN) has emerged, which switches the end-to-end communication paradigm of today's IP-based Internet to a content-centric communication. Instead of using a host address to retrieve information, ICN directly seeks for the user requested data. As a result, the requested data can be sourced from anywhere, eliminating the need of establishing a connection with the host every time. In addition, ICN allows the nodes in a network to cache data. This caching feature allows user to retrieve requested data from a nearby node. ICN focuses on establishing a future Internet architecture that is faster, more reliable, and more secured than the existing IP-based Internet~\cite{ICN2015}.

% There are a number of ongoing projects under the umbrella of ICN, such as content-centric network (CCN), named data network (NDN), data oriented network architecture (DONA) proposed by Berkeley University~\cite{DONA}, network of information (4WARD-NetInf)~\cite{4WARD}, etc. In addition, some researchers are trying to combine ICN with SDN to propose a novel data-driven future Internet architecture. Others are exploring the possibility of implementing AI in ICN~\cite{ai_icn2019}.

At the same time, NDN is the most mature information-centric networking architecture~\cite{zhang2014named}. NDN is destined to replace the current IP-based Internet architecture with a data-centric communication model. In a traditional IP network, a user's request contains the information of the desired host, and a connection is established between the host and the user for data exchange. Contrary to that, a user directly requests the desired data in NDN. %In other words, the requested data is directly received by the network nodes. As long as the data is authentic and valid, NDN does not care about the origin of the data. As such, 
In NDN, each piece of data is identified through a \textit{name} and carries a cryptographic signature generated by the producer of the data. %Names in NDN are hierarchical, which means it indicates the path to the host of the data. 
Each NDN router stores the information necessary to forward requests for data in its Forward Information Base (FIB). NDN routers also cache copies of requested data in their Content Store (CS). Requested data can be satisfied with data cached in CS, reducing the incurred latency and increasing efficiency. %Adaptive forwarding is one of the key features of NDN. 
Finally, NDN routers aggregate requests for the same data in their Pending Interest Table (PIT), which prevents overloading the network with multiple requests for the same data.

%In addition, NDN provides better security by encrypting the data directly, instead of encrypting the connection between the host and user.

\subsection{Related Work}

F-RAN has attracted a lot of attention in recent years. Since the key feature of F-RAN is edge caching, most of prior works are focused on optimized content placement and delivery scheduling. Kavena \etal~proposed an optimal energy-aware scheduling algorithm using network coding to offload popular contents from the BBU to F-UEs~\cite{kavena2020}. %The proposed scheduling algorithm encodes and transmits file within the given energy constrain of the F-UE using half-duplex transmission to minimize BBU resource usage. 
Shnaiwer \etal~proposed a cache offloading scheme based on opportunistic network coding for macrocell base stations using femtocache in F-RAN~\cite{femto2018}. The combination of F-RAN with heteregeneous wireless technologies, such as LTE and WiFi, was also explored to increase throughput~\cite{8701695}. Hu \etal~reduced content response latency by adopting a caching design for enhanced remote radio heads based on a geographic distribution~\cite{errh2020}.

Several applications have shown improved performance with the help of NDN~\cite{v2indn}. %Prospective ICN-based cellular network solutions are also emerging. 
Lei \etal~proposed a probability-based multipath forwarding using NDN for 5G~\cite{ndn5g2018}. %The work incorporated network coding in order to increase the throughput of internet of things (IoT) devices in the 5G network. 
Liao \etal~, motivated by the upcoming surge in virtual and augmented reality, proposed a large-scale content distribution model for 6G using NDN~\cite{vrar6gicn2020}. The application of NDN in fog networks have been also investigated. Hua \etal~ presented a fog caching design scheme in NDN~\cite{fogicniot2020}. %The work incorporated three designing schemes: a fog cluster-based scheme to cache content closer to the device, a near-path approach leverage nearby caching nodes, and a reactive/proactive caching scheme to avoid network congestion. 
Liang \etal~explored the possibility of combining NDN and wireless network virtualization~\cite{ICNVirtual5G}. The work by Zhang \etal~presented a NDN-enabled 5G architecture based on the 3GPP standards~\cite{edgecache5g}. %However, the implementation was only within the RAN, and packets leaving the core network (5G Core) must be converted to an IP protocol.

%As for the optimal cache allocation in general ICN networks, progress has been made in recent years. 
Furthermore, approaches to optimize in-network data caching in NDN have been explored. ProbCache in~\cite{probCache} was introduced as a probability-based in-network caching algorithm. Zhang \etal~in~\cite{edgecache5g} employed an auction-based caching strategy to reduce data access delay and increase efficiency in content delivery. H2NDN in~\cite{h2ndn2020} was proposed with a frequency-based cache allocation scheme in order to move popular cached data closer and less popular data further from users. Finally, an optimal cache budget distribution for NDN networks was proposed by Montazeri and Makaroff~\cite{arrivalratecache}.

\subsection{Our Motivation and Contributions}
% NDN is a potential candidate for the future of Internet, which might replace the existing IP-based Internet architecture. 

In this paper, our motivation is to study the interplay between NDN and F-RAN. This can be advantageous, as we further discuss below.
%However, this has not been studied yet. In our study, we try to explore the possibility of NDN-enabled F-RAN, along with a better cache distribution strategy to optimally utilize the overall cache resources in F-RAN.
A key feature of the F-RAN edge devices is that they can have individual storage capabilities to cache data. %The F-RAN architecture includes various cache enabled edge devices like eRRH, macrocell base station, femtocache, and F-UE.
Since F-RAN reduces latency and fronthaul usage by bringing popular data closer to users, it is crucial to have strategies and algorithms to identify popular data and place it on edge devices. These are challenging tasks that we need to carry out in order to have a feasible system. Incorporating NDN into the F-RAN architecture allows us to take advantage of the existing cache resources available on edge devices to realize NDN in-network caching. At the same time, F-RAN will no longer need to perform data caching at the application layer--caching will be performed by NDN directly at the network layer. 

%This has an intrinsic benefit; there will be no need to identify popular contents and place them in the caches. Moreover, it will be faster and more efficient since the cache in NDN is in-network. Using the default cache placement and replacement strategies of the NDN architectures is not the most efficient approach. In this work, a more efficient cache distribution strategy for F-RAN is proposed. However, because NDN-enabled F-RAN is not yet studied, some design changes to the components of F-RAN are required to enable the NDN protocol, which is also presented in this paper.

%F-RAN may benefit the most if it is incorporated with NDN. The edge caching feature can be easily utilized by NDN to store in-network cache. At first, we need client and server applications that uses NDN as the communication protocol. The F-UE, F-AP, BBU pool, and the core network should also be able to process and route NDN packets. The proposed modification of these components are on the protocol stack. 

To realize an NDN-enabled F-RAN architecture, in this paper, we augment the protocol stack of the F-UE and F-AP, so that they are able to understand the semantics of NDN packet. In addition, we propose a data caching strategy to improve cache usage. Our contributions are the following:

\begin{itemize}
    \item We propose an NDN-enabled F-RAN architecture through modified designs of the F-UE and F-AP protocol stack to incorporate NDN into F-RAN.
    
    \item We present a cache distribution problem along with an optimization function to effectively utilize the available cache resources in the network. Additionally, we propose a cache replacement algorithm for the NDN-enabled F-RAN components to improve cache usage.
    
    \item Through a simulation study, we compare our data caching strategy with baseline strategies in terms of resource utilization, cache hits, and fronthaul channel usage.
    
\end{itemize}

%We believe the aforementioned contributions will help in the advancements of future radio access networks.

% \subsection{Paper Organization}
% The rest of the paper is arranged as follows. Section~\ref{sec:ndn-fran} discusses the fundamental NDN principals and how NDN can be enabled in the existing design of F-UE, F-AP, and BBU pool. Section~\ref{sec:sys_model} demonstrates the problem with default caching strategies and how it can be further optimize to utilize the total available cache space. An algorithm is also presented in this section for the NDN-enabled F-RAN components. Section~\ref{sec:result} compares the efficiency of the proposed caching strategy by comparing it with the default ones in different cases. Finally, conclusions are drawn in Section~\ref{sec:con} with some future research directions.

%% file: 3-ndn_fran.tex
\section{Enabling NDN in F-RAN}\label{sec:ndn-fran}

Modifications are needed for the F-RAN architecture to support NDN. These modifications are discussed in this section. %The core design principles of F are mostly preserved in our proposed designs.

\subsection{NDN Principles}

There are two types of packets in NDN: \textit{Interest packets} and \textit{Data packets}. An \textit{Interest packet} contains the name of the Data requested by a consumer (user), and a \textit{Data packet} contains the requested content along with some meta information and a signature. In NDN, data names are hierarchical, which allow routers to forward packets to their next hop(s) accordingly. Each NDN router maintains three data structures for its operation: FIB, PIT, and CS. %FIB maintains a table of routing information based on names. When a router is awaiting a \textit{Data packet} in response to an Interest, every subsequent Interest for the same data is kept on-hold (aggregated) in PIT. A copy of every \textit{Data packet} received by a router is cached in its CS.

When an \textit{Interest packet} arrives at a router, the router first checks if the requested \textit{Data packet} is available in CS. If it is available, the router satisfies the Interest with the data from its CS. Otherwise, the router checks if the requested \textit{Data packet} has already been requested by another consumer. If an Interest for the requested data is already in PIT, the router aggregates the new Interest. Otherwise, the router forwards the \textit{Interest packet} to its next hop based on the name-based routing information available in FIB and creates a new PIT entry (stateful forwarding plane)~\cite{al2022reservoir}. When the requested data is received by a router as a \textit{Data packet}, it is cached in its CS for future requests, and all the aggregated Interests in PIT are satisfied with the \textit{Data packet}. %NDN routers are stateful, which allows them to see what content a consumer is asking for and act upon it. In contrast, the routers only know the destination of the packet in an IP-based network; thus, it must be satisfied by the server sitting in the destination.

\begin{figure}[!t]
\centering
\includegraphics[width=\linewidth]{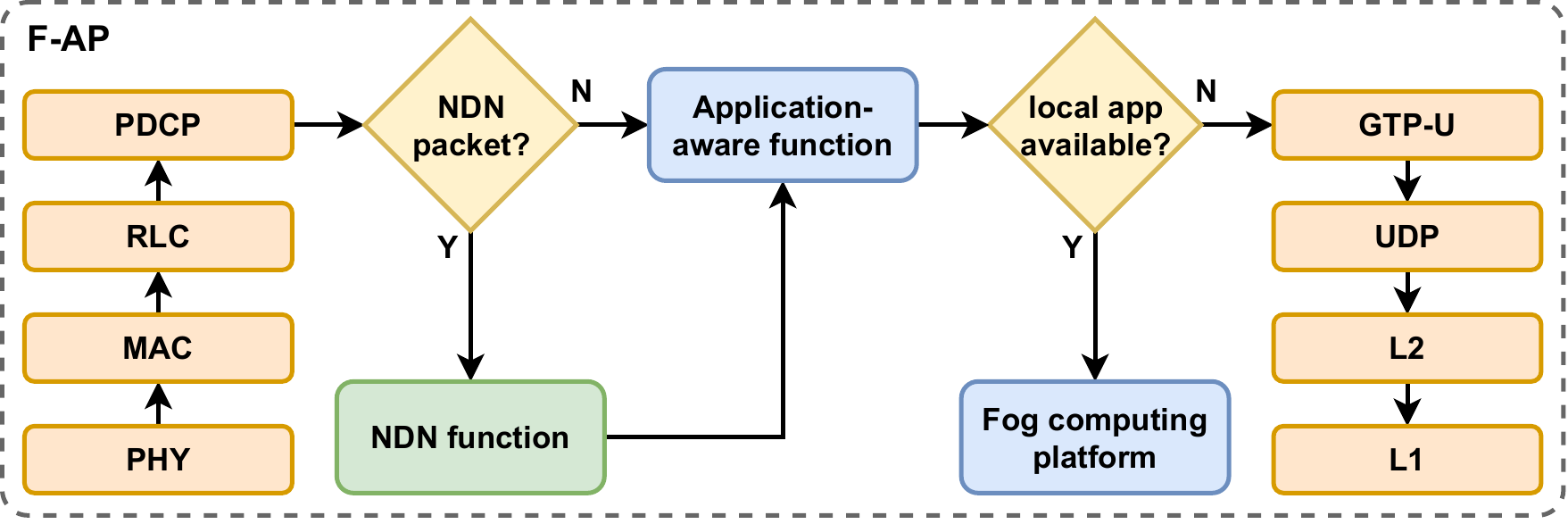}
\includegraphics[width=\linewidth]{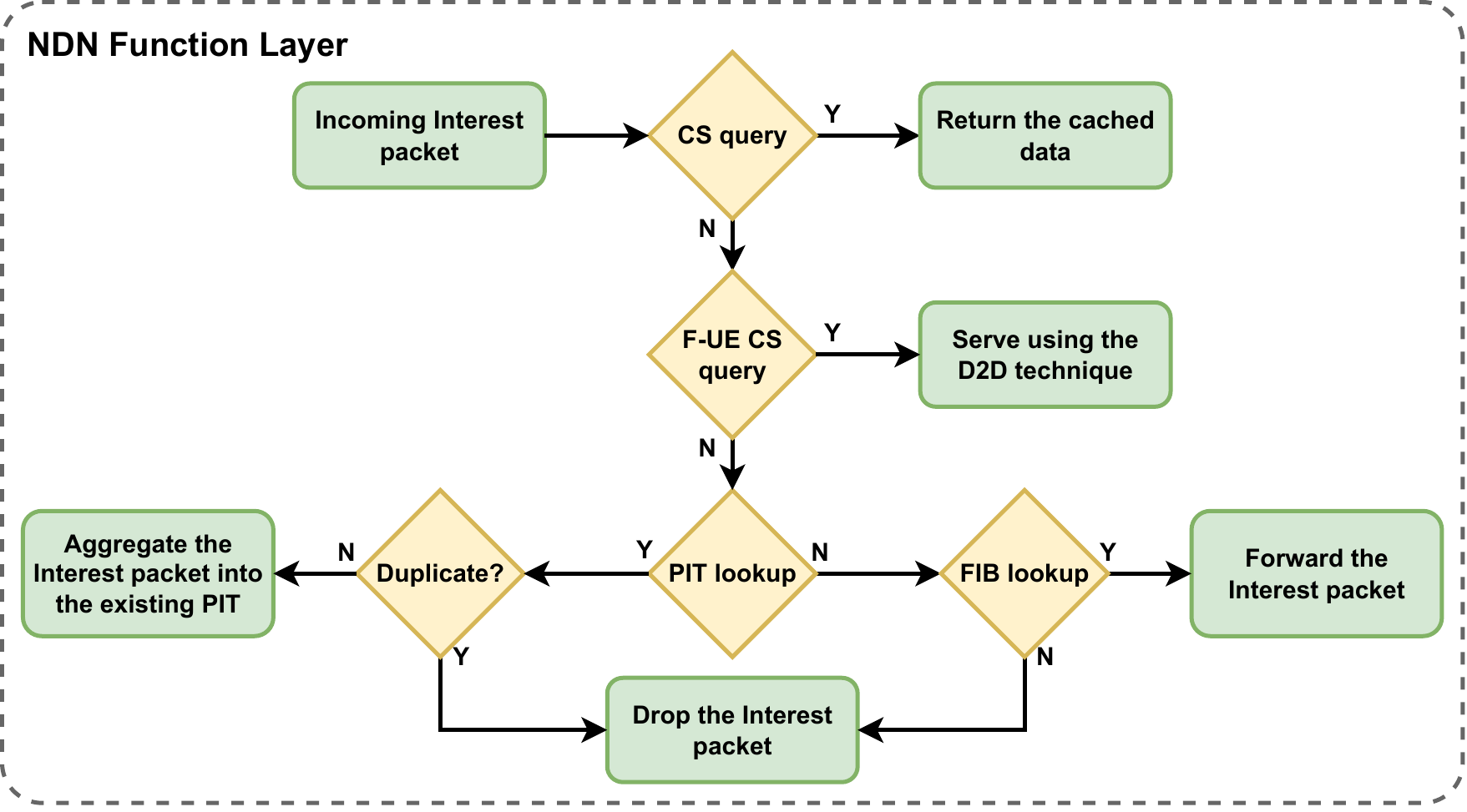}
\caption{NDN-enabled F-AP protocol stack.}
\vspace{-0.3cm}
\label{fig:fran_stack}
\vspace{-0.3cm}
\end{figure}

% \begin{figure}[!t]
% \centering
% \includegraphics[width=\linewidth]{figure/NDN Function Layer.pdf}
% \caption{NDN function layer.}
% \label{fig:fran_ndn_fn}
% \vspace{-.5cm}
% \end{figure}

\subsection{NDN-Enabled F-AP}

The original design of an F-AP for F-RAN has been proposed by Peng \etal~\cite{FRANBookCh9}. The F-AP is based on a four layer protocol stack with additional functions for fog computation. 3GPP standards are preserved in the design. The physical (PHY), Medium Access Control (MAC), and Radio Link Control (RLC) layers are the same as C-RAN with the functionality of coding/modulation, mapping logical/transport channel, and segmentation/reassembly respectively. The Packet Data Convergence Protocol (PDCP) layer (responsible for IP header compression) sends packets to an application-aware function. The function extracts the IP address and port number of the requested application. A table of IP addresses and ports of the available fog applications is maintained, where the extracted IP and port number is matched. If the requested application is available, the function notifies the fog computing platform, initiating the requested application. The newly created application's IP address and port number are forwarded to the user. %The user then requests the application with the local IP address and port. 
If the application is not available, the packet is forwarded to the GPRS Tunneling Protocol (GTP-U) layer.

To enable NDN support for the F-AP design, an NDN function layer supporting the NDN protocol is proposed in Figure~\ref{fig:fran_stack}. The PHY, MAC, and RLC layers do not need modifications as these layer do not deal with the IP addresses. However, the PDCP layer needs to be updated to support both IP and NDN packets. When the PDCP layer forwards an NDN packet, it will be processed at the NDN function layer. The CS in the F-AP maintains the original cache and the information of the cache of its F-UEs. The function looks at the name of an \textit{Interest packet} processed by the PDCP. If the requested data is available in the F-AP's CS, it is returned right away. Otherwise, the function will check if one of the nearby F-UEs (within the consumer's D2D communication region) has the requested data. If an F-UE has the data cached in its CS, D2D communication is established, allowing the F-UE to serve the data. The functionalities of PIT and FIB remain the same. The application-aware function needs to be updated as well to support NDN packets. The updated function will maintain an additional table with the names of the locally available applications. If an \textit{Interest packet}'s name is found in the table, the fog computing platform will be notified. An NDN packet can be forwarded to the serving gateway using GTP-U and UDP/IP without modifications. Our proposed NDN-enabled F-AP design is able to handle both IP and NDN traffic.

\begin{figure}[!t]
\centering
\includegraphics[width=100pt]{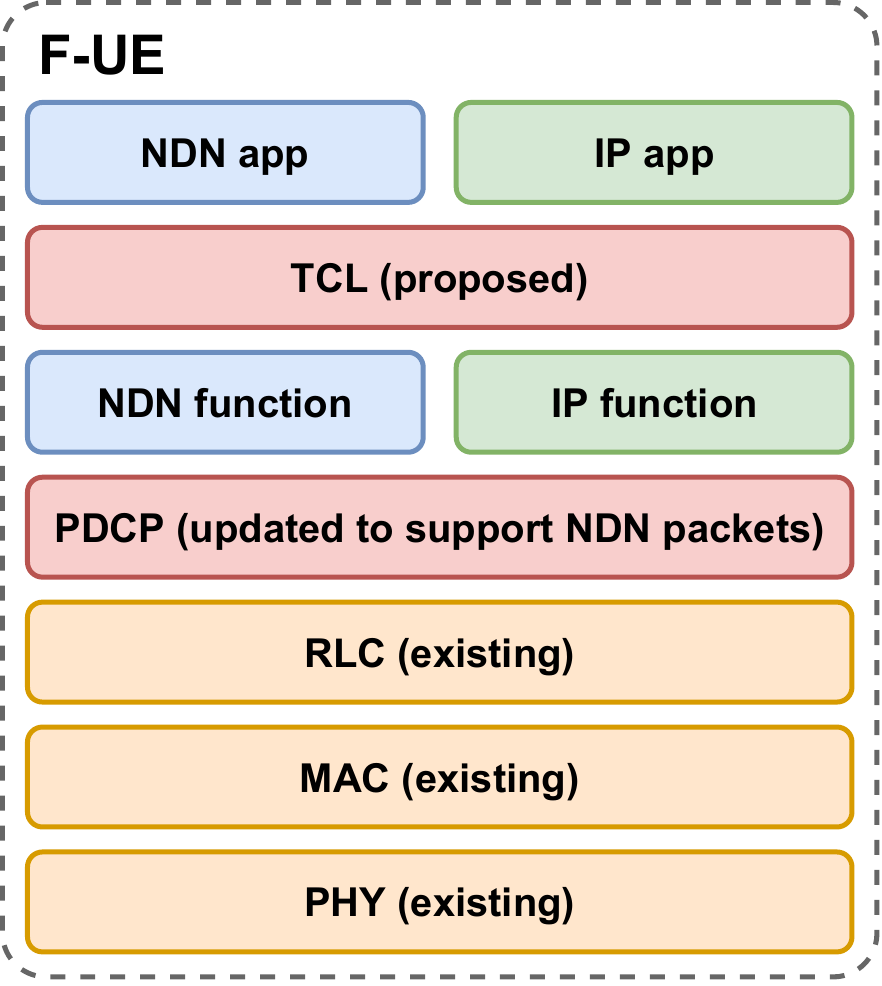}
\vspace{-.2cm}
\caption{Dual stack protocol design for an NDN-enabled F-UE.}
\label{fig:fue_stack}
\vspace{-.3cm}
\end{figure}

\subsection{NDN-Enabled F-UE and BBU Pool}

To enable NDN support for F-UEs, we present a dual stack protocol design (Figure~\ref{fig:fue_stack}). This protocol design ensures support for both NDN and IP. It will be up to the application to use either NDN or IP as the communication protocol. We propose an additional Transport Convergence Layer (TCL), which is responsible for sending NDN packets over an IP network if a radio access network does not support native NDN. The NDN packets can also be tunneled through IP over a network that does not have native NDN support using the NDN forwarding daemon~\cite{nfd-dev}. %The purpose of the proposed NDN function layer is the same as the one in F-AP, with an exception of no F-UE CS query for D2D communication. The RLC, MAC, and PHY layer does not need any modification as they are not IP-aware.

The design of the BBU pool also requires changes to incorporate NDN. Our design of the BBU pool incorporates a core NDN component, which realizes the NDN protocol stack and is responsible for handling NDN traffic. Our design also incorporates a protocol conversion component, so that the BBU pool can communicate with the core network in cases that NDN is not supported by the core network. This is achieved by encapsulating NDN packets into IP packets.

%% file: 4-sys_model.tex
\section{System Model}
\label{sec:sys_model}

% The proposed NDN enabled F-RAN components in the previous section can be employed without further modification. 
NDN allows F-RAN to store data directly at the network layer. The CS of each network node is automatically populated based on the data requested by users. %However, the default cache placement and replacement algorithms are not efficient. 
In this section, we elaborate on the data caching problem and we present a cache replacement strategy to improve caching effectiveness.

%discusses the need of a better cache distribution strategy to make proper use of the available cache.

\begin{figure}[!t]
\centering
\includegraphics[width=2.5in]{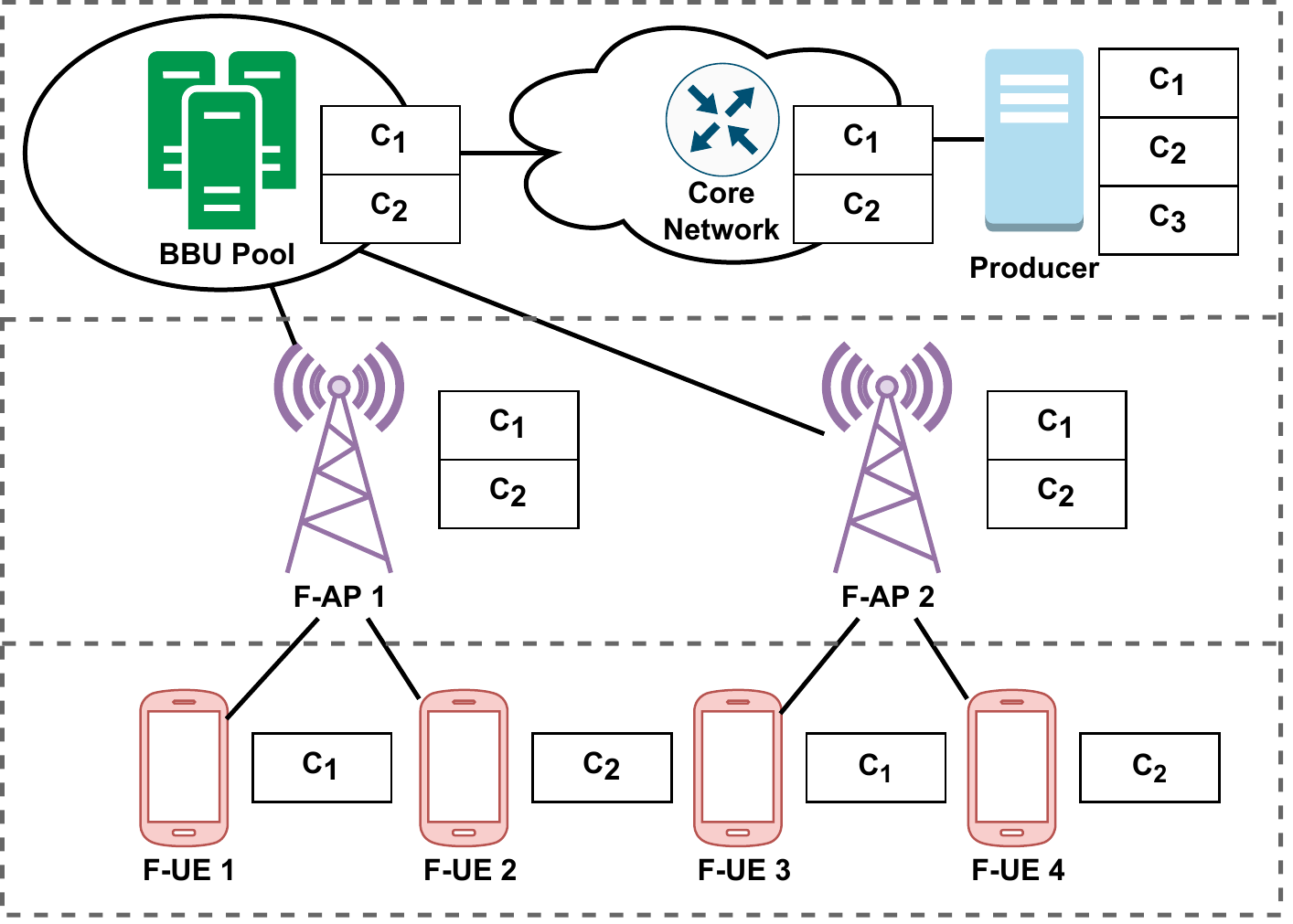}
\caption{Demonstration of the cache allocation problem with naive caching strategies in NDN.}
\label{fig:fran_lce}
\vspace{-0.5cm}
\end{figure}

\subsection{Data Caching Problem}

In NDN, data caching strategies, such as Leave Copy Everywhere (LCE) for cache placement and First-In-First-Out (FIFO) for cache replacement, have been commonly used. Such strategies may not be suitable for F-RAN as every node in the network may cache the same data. To demonstrate the problem, a simple scenario is presented in Figure~\ref{fig:fran_lce}. In this scenario, there is a producer generating contents: $C_1$, $C_2$, and $C_3$. A BBU from the BBU pool is serving F-AP 1 and F-AP 2. F-UE 1 and F-UE 2 are served by F-AP 1, and F-UE 3 and F-UE 4 are served by F-AP 2. Initially, every node's CS is empty. In this scenario, both the BBU pool and F-APs can cache two contents, and each F-UE can cache one content in CS. The contents are requested in phases as described below.

First, F-UE 1 requests $C_1$ from the producer. F-AP 1 forwards the \textit{Interest packet} to the serving BBU pool through the fronthaul channel and the BBU forwards the \textit{Interest packet} to the core network via the backhaul channel, which eventually reaches the producer, and the requested \textit{Data packet} is sent. Because of the LCE placement policy, the BBU, F-AP 1, and F-UE 1 will all store a copy of $C_1$ in their CS. Next, F-UE 2 requests $C_2$. As a result, the CS in the BBU pool and F-AP 1 is now filled with $C_1$ and $C_2$. Subsequently, F-UE 3 and F-UE 4 request $C_1$ and $C_2$ respectively. The BBU pool has both $C_1$ and $C_2$ available in its CS and it will serve both. However, F-AP 2 also has $C_1$ and $C_2$ in its CS. The BBU pool and the F-APs store the same content, which underutilizes the total available caching resources. Next, F-UE 1 and F-UE 2 request $C_3$ from the producer. Since every CS in the network is full, cache replacement is necessary to incorporate $C_3$. If FIFO is used, $C_1$ will be replaced by $C_3$ in the BBU pool, F-AP 1, F-UE 1, and F-UE 2. %not being able to effectively utilize caches. Clearly there is room for improvements. 
%For example, 
Instead of replacing $C_1$ in the CS of both the BBU pool and F-AP 1, we could replace only F-AP 1's cache with $C_3$. This would have utilized cache resources more effectively, since if F-UE 1 or F-UE 2 were to request $C_1$ again, the BBU pool could serve it.

% An optimal cache distribution strategy for F-RAN is proposed in this paper to maximize the use available total storage in each node. BBUs generally have the most cache storage, and as we go down the hierarchy of the network, the cache storage reduces (i.e., an F-AP have less cache than a BBU pool). While being the fastest, an F-UE stores the least amount of cache in F-RAN. No hop is needed when the content is cached at the user's F-UE; thus, it can be served immediately. If the a content is served by an F-AP, it requires one hop, and no fronthaul channel is used. The BBU pool servers a content from its cache via the fronthaul channel, which takes two hops. Lastly, if a content comes from the core network, it takes three hops~\cite{FRANBookCh7}. Based on these circumstances, more popular caches should be distributed closer to the F-UEs, while less popular caches should be moved to the BBU pool.  

\begin{table}[!t]
\caption{List of notations.}
\label{tab:notation}
\centering
\begin{tabular}[width=\linewidth]{cl}
\hline
Notation & Definition\\
\hline
$\mathcal{D}$ & Set of all \textit{Data packet}s.\\
$\mathcal{N}$ & Set of all nodes (BBUs, F-APs, and F-UEs).\\
$\mathcal{U}$ & Set of all F-UEs where $\mathcal{U}\subset\mathcal{N}$ \\
$\mathcal{A}$ & Set of all F-APs and BBUs where $\mathcal{A}\subset\mathcal{N}$.\\
$\mathcal{C}_n$ & Set of all children of $n \in \mathcal{N}$.\\
$\lambda_d^n$ & Request rate of $d \in \mathcal{D}$ at $n \in \mathcal{N}$.\\
$x_d^n$ & Binary decision variable for cache availability\\&  of $d \in \mathcal{D}$ at $n \in \mathcal{N}$.\\
$h_n$ & Hop distance from the core network to $n \in \mathcal{N}$.\\
$\mathit{CS}_n$ & CS size of $n \in \mathcal{N}$.\\
\hline
\end{tabular}
\vspace{-.5cm}
\end{table}

\subsection{Proposed Caching Strategy}

First of all, popular cached contents need to be identified, which can be done by analyzing the request rate of a content piece at a particular node. In other words, if the rate of requesting a specific \textit{Data packet} is higher at a node, the node should hold on to that cached content for a longer period of time. For example, more popular contents based on the request rate will be served by the F-AP and less popular contents will be served from the BBU pool. Since the BBU pool will only receive the \textit{Interest packet}s not satisfied by F-APs, requests for less popular caches should aggregate there. Subsequently, the cache will be distributed among the nodes in the network based on the request rate of the \textit{Data packet}s at each node. To improve cache distribution efficiency, cached contents with a high request rate should be cached further away from the core network (i.e, closer to users). This can be achieved by optimizing the following objective function (notations are defined in Table \ref{tab:notation}):
\begin{align}
    \label{eq:obj_fn}
    \max_{\lambda,x,d} \: &\sum_{\forall n \in \mathcal{N}} \sum_{\forall d \in \mathcal{D}} \lambda_d^n x_d^n h_d^n\\
% \end{equation}
% \begin{equation}
    \label{eq:st}
    \textnormal{s.t.} \: &\sum_{\forall d \in \mathcal{D}} x_d^n \leq \mathit{CS}_n \quad \forall n \in \mathcal{N},
\end{align}

This indicates that caching data with greater request rates ($\lambda$) further away from the core network based on number of hops ($h$) should be maximized within the limit of $\mathit{CS}$ at each node ($n$). At an F-UE, the request rate of a data piece $d$ is defined as the number of \textit{Data packets} not satisfied by the consumer F-UE; whereas the request rates at the F-APs and BBU pool are the total number of unsatisfied \textit{Interest packets} coming from its children within a given period of time. The request rate for each \textit{Data packet} is calculated using the following equation:
\begin{equation}
\label{eq:rate}
\lambda_d^n = \begin{cases}
\sum_{\exists d \in \mathcal{D}} 1 - x_d^n &\forall n \in \mathcal{U}\\
\sum_{\forall c \in \mathcal{C}_n} \lambda_d^c(1 - x_d^c) &\forall n \in \mathcal{A},\\
\end{cases}
\end{equation}
where $x_d^n$ is a binary decision variable, which indicates the availability of data $d$ at node $n$. As such, it can be defined as follows:
\begin{equation}
\label{eq:bin_var}
    x_d^n \in \{0, 1\} \quad \quad d \in \mathcal{D}, \quad n \in \mathcal{N}.
\end{equation}
The problem in \eqref{eq:obj_fn} is NP-hard as it can be reduced to a zero-one linear programming problem. The proof is presented in Appendix~\ref{sec:apndx1}.

\subsection{Data Caching Algorithm}

The request rate for each \textit{Data packet} $d$ is calculated using \eqref{eq:rate}. However, the request rate $\lambda$ is subject to change at any time; thus, a node should update the request rates after $\tau$ seconds. To make the changes in a more granular rate, a weighted average approach is taken into account based on the following equation:
\begin{equation}
    \label{eq:update_rate}
    \lambda_{updated} = \frac{\alpha \lambda_{new} + \beta \lambda_{old}}{\alpha + \beta},
\end{equation}
where $\alpha$ is the weight of the newly calculated request rate after $\tau$ seconds, and $\beta$ is the weight of the existing request rate. Algorithm~\ref{algo} presents the cache distribution algorithm, which runs on each individual node in F-RAN. The algorithm is based on the principals of NDN along with our proposed caching strategy. When a node receives \textit{Interest packets} from a child node (line \ref{algo1:interest_handling}), it checks $\mathit{CS}, \mathit{PIT}$, and $\mathit{FIB}$ to forward the \textit{Interest packet} according to the NDN protocol. When the algorithm receives the corresponding \textit{Data packet}, it satisfies the pending requests (line \ref{algo1:data_arrive}). To cache the \textit{Data packet}, it first checks if there is any space available in $\mathit{CS}$ (line \ref{algo1:check_cs}). Otherwise, it finds the \textit{Data packet} with minimal $\lambda$ and hops and compares it with the received \textit{Data packet} (line \ref{algo1:find_min}). If the existing \textit{Data packet} has lower $\lambda$ and hops, it is replaced by the new \textit{Data packet} (line \ref{algo1:replace_cs}). After $\tau$ seconds the algorithm updates the request rates ($\lambda$) of each \textit{Data packet} using \eqref{eq:update_rate} (line \ref{algo1:update_rate}). The complexity of the algorithm is $\mathcal{O}(m+k)$, where $m$ is the size of the $\mathit{CS}$ and $k$ is the size of $\mathcal{D}$.

%\vspace{1cm}

%\vspace{-0.5cm}

%% file: 5-results.tex
\section{Simulation Results and Discussion}
\label{sec:result}

We conduct a simulation study to evaluate the the proposed caching strategy, which we compare with two baseline caching algorithms, FIFO and Least Recently Used (LRU). Our simulation topology contained a central BBU pool, five F-APs, and a range of F-UEs (ranging from five to thirty). The central BBU pool served five F-APs, and each F-AP served one to six F-UEs. Each F-UE generated Interest packets %per second for fifty seconds from the pool of twenty (
following a Zipf distribution. Simulations were done with D2D communication enabled and disabled, assuming that each F-UE under an F-AP was within the range of D2D communication.

\begin{figure*}
  \centering 
  \subfloat[\label{fig:avg_hops}]{\includegraphics[width=6cm]{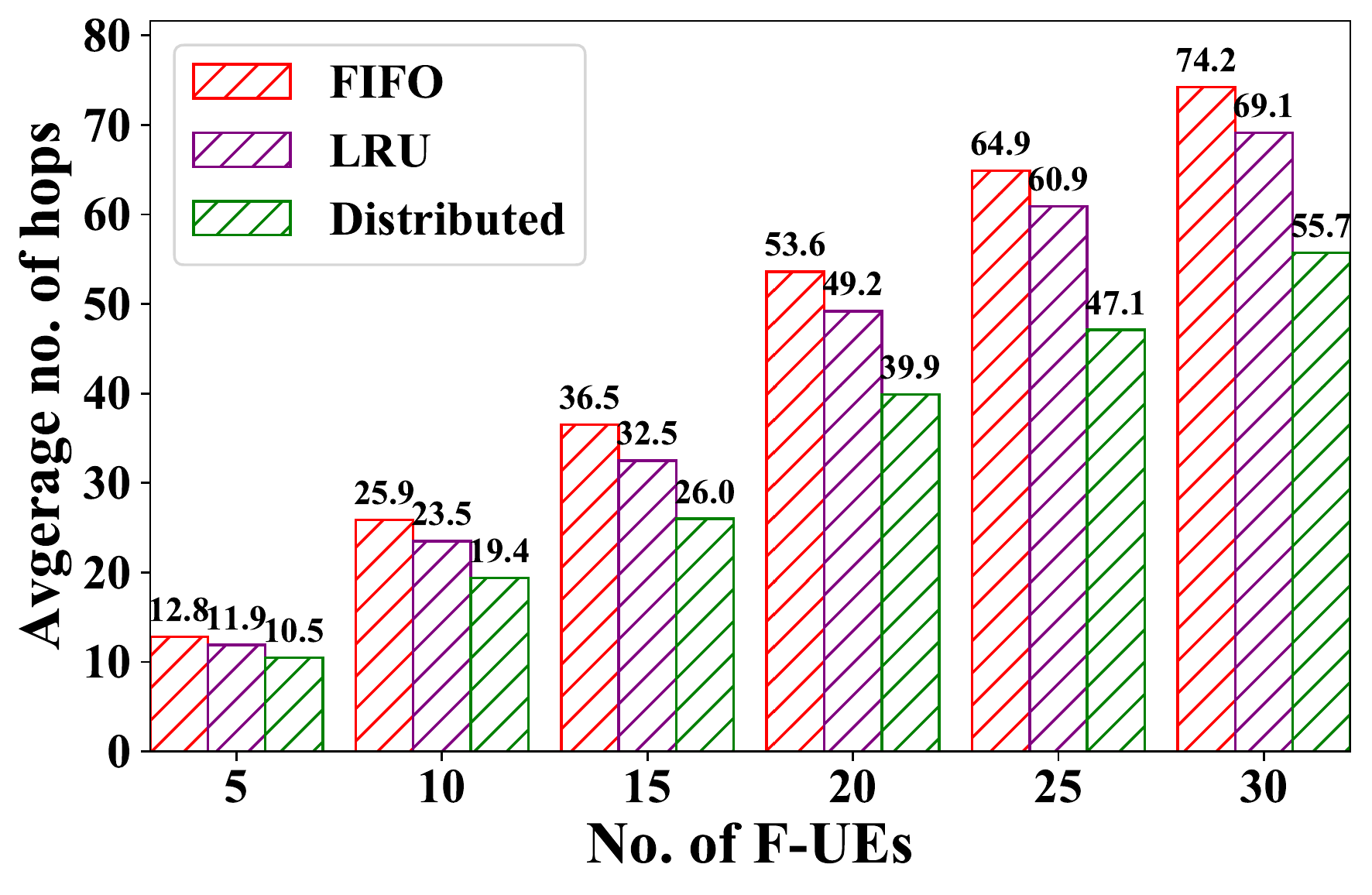}} \hfill
  \subfloat[\label{fig:avg_hops_d2d}]{\includegraphics[width=6cm]{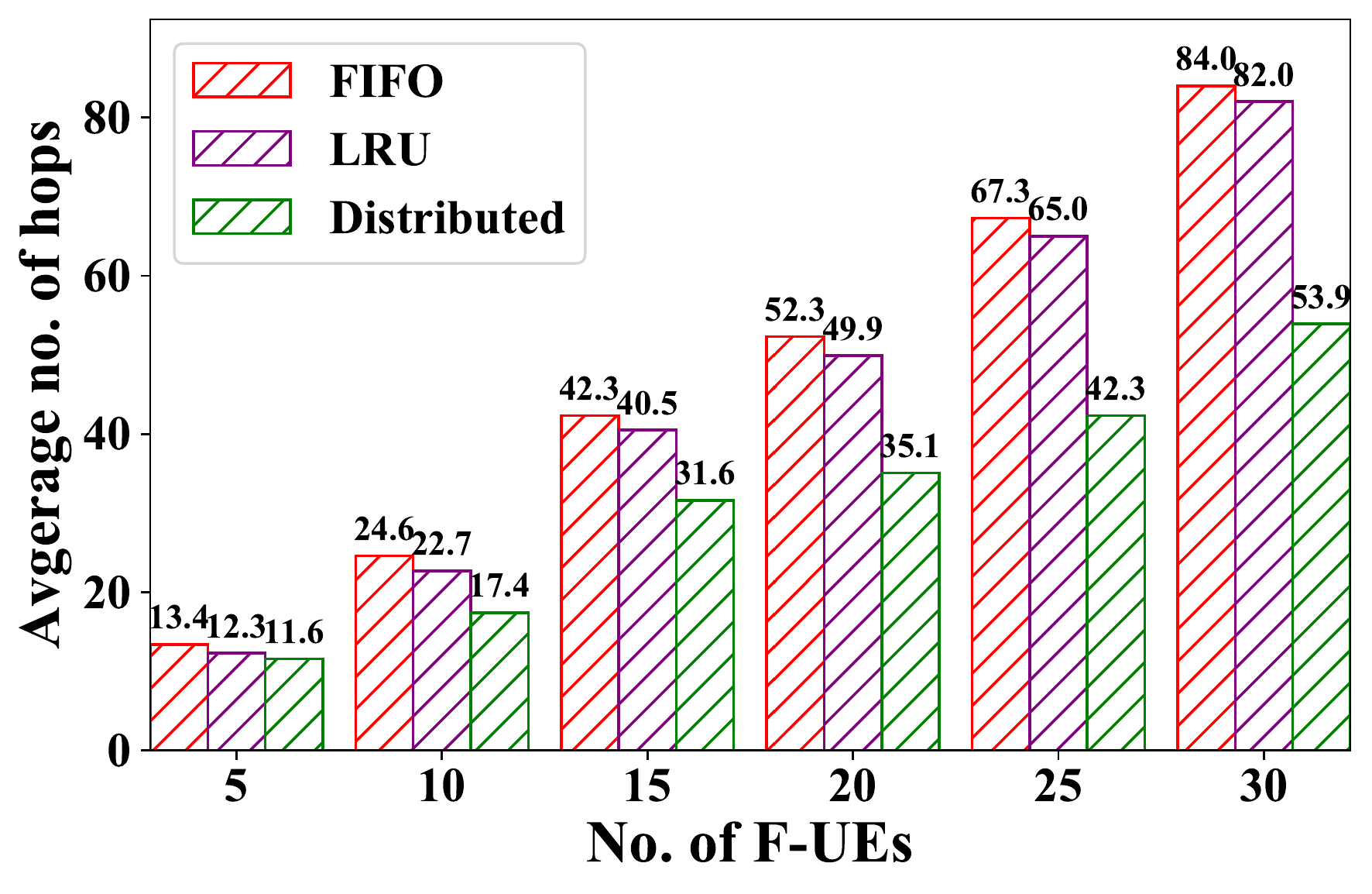}} \hfill
  \subfloat[\label{fig:cache_hit}]{\includegraphics[width=6cm]{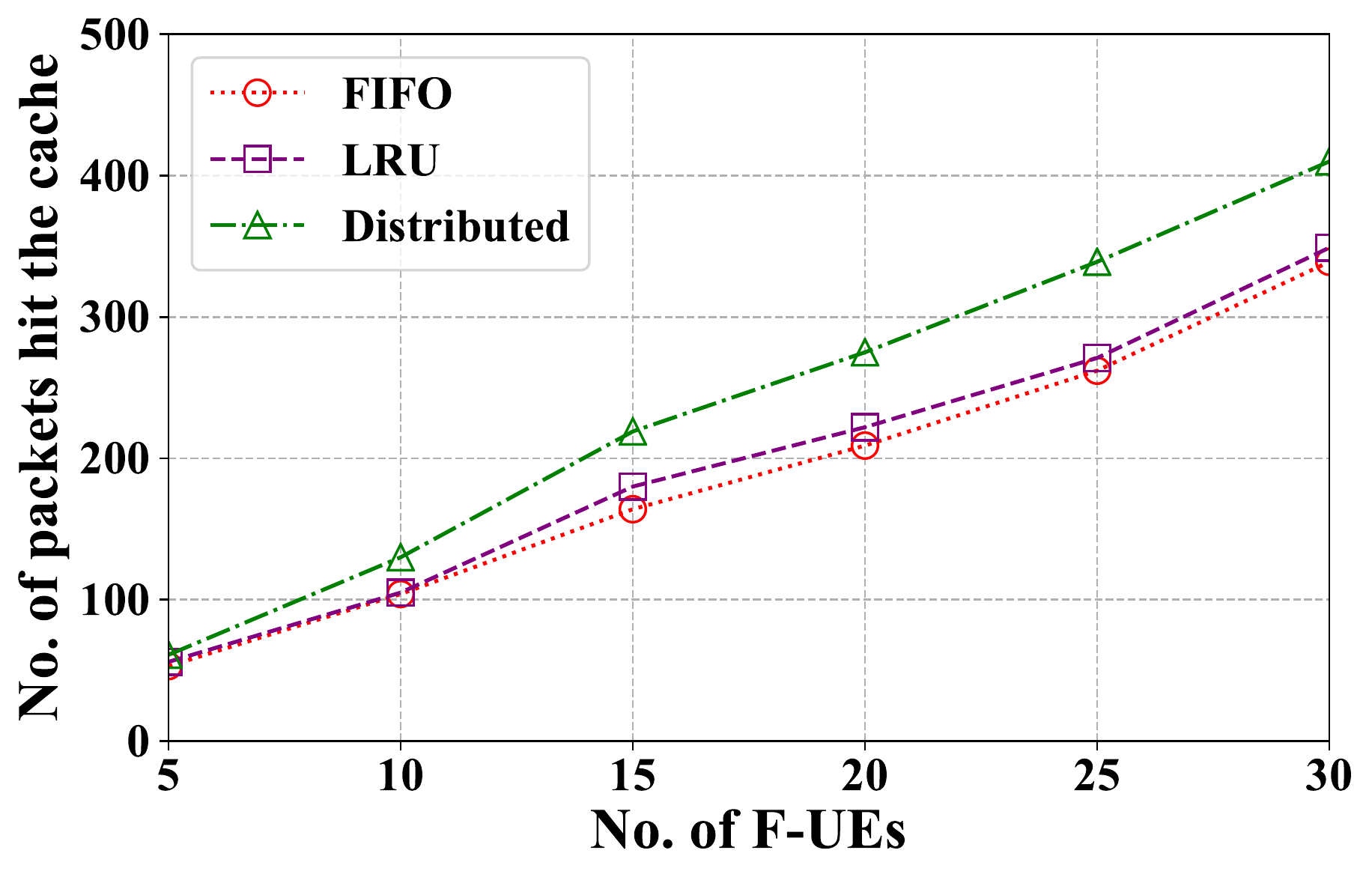}}
  \vspace{-0.4cm}
  \subfloat[\label{fig:cache_hit_d2d}]{\includegraphics[width=6cm]{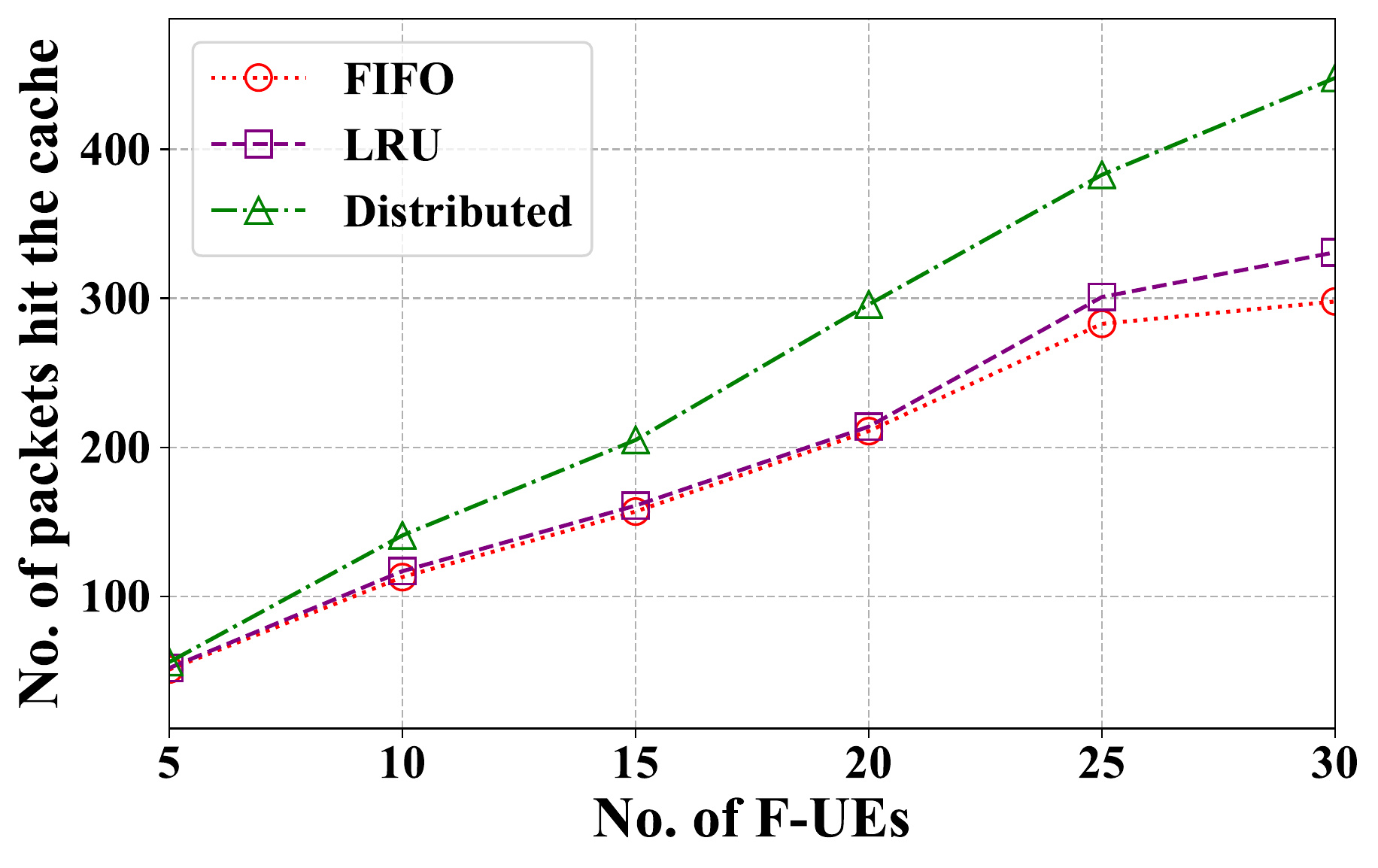}} \hfill
  \subfloat[\label{fig:fh}]{\includegraphics[width=6cm]{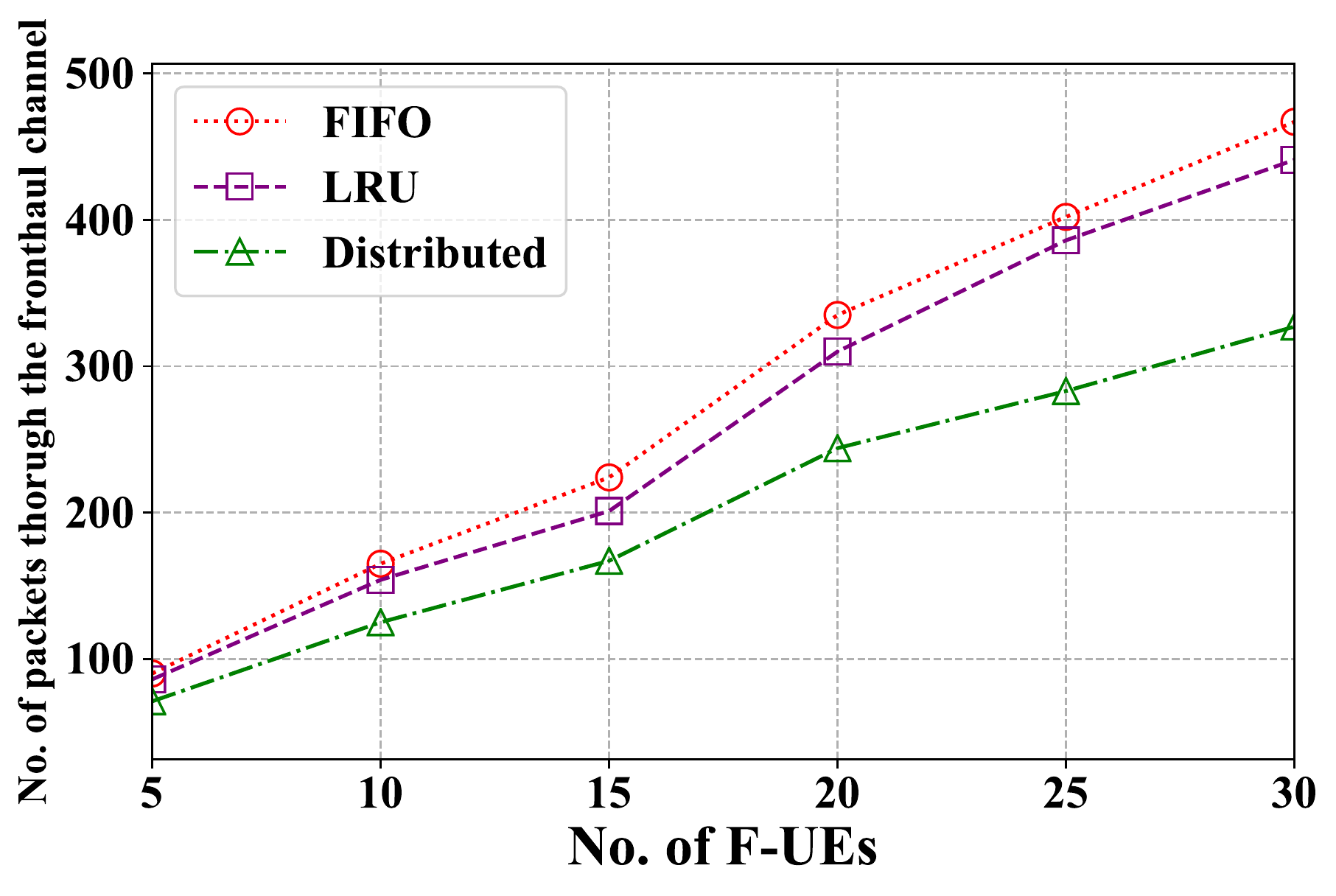}} \hfill
  \subfloat[\label{fig:fh_d2d}]{\includegraphics[width=6cm]{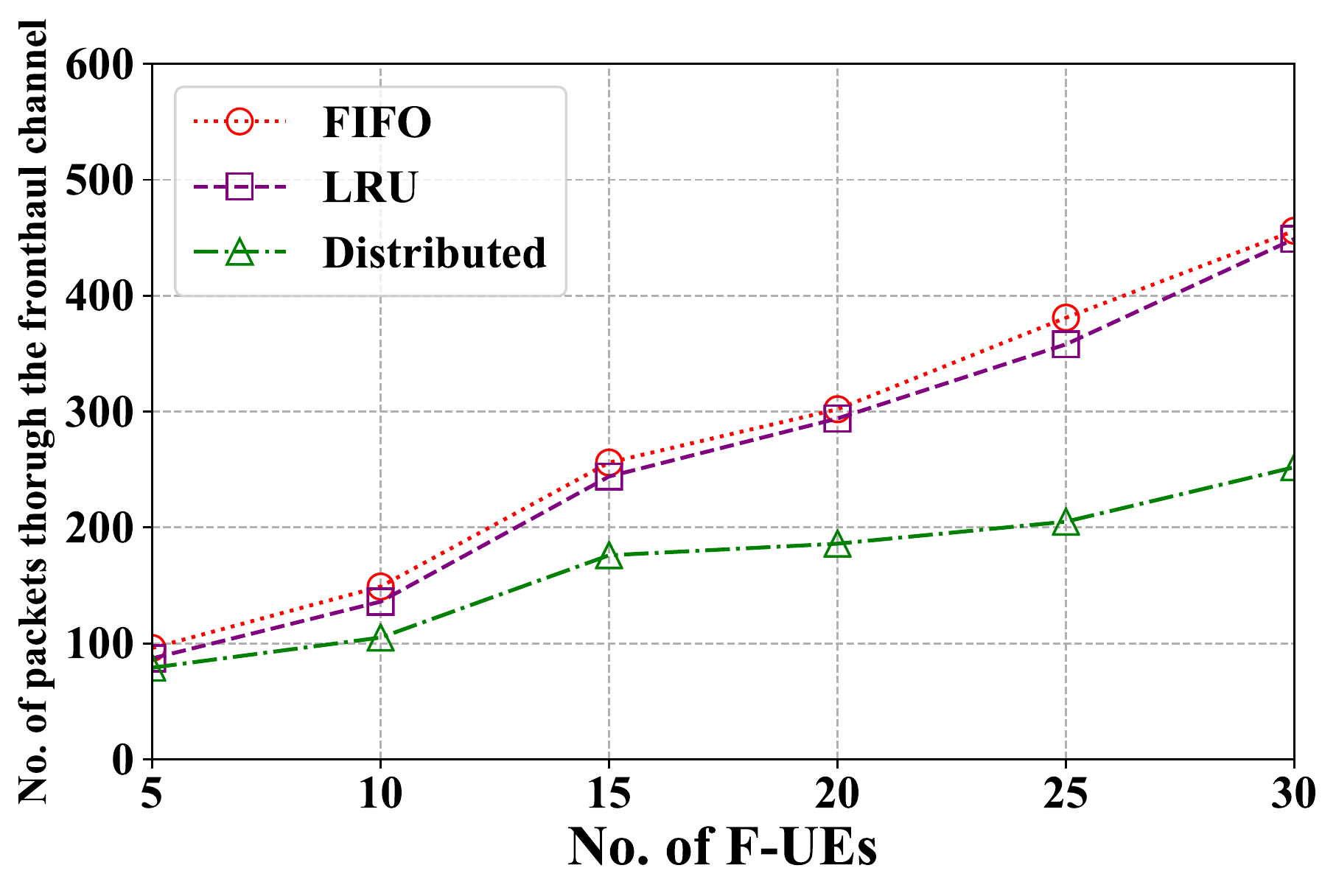}}
  \vspace{-0.2cm}
  \caption{Simulation results of average hops (a) without D2D and (b) with D2D communication, in-network cache hits (c) without D2D and (d) with D2D communication, and fronthaul usage (e) without D2D and (f) with D2D communication.}
  \label{fig:results}
  \vspace{-0.5cm}
\end{figure*}

%\subsection{Average Hops}

\noindent \textbf{Average number of hops:} Figures~\ref{fig:avg_hops} and~\ref{fig:avg_hops_d2d} present the average number of hops used by F-UEs without and with using D2D communication respectively. As the number of F-UEs increases, the number of requests and the required hops increase as well. Fewer hops indicate that an Interest was satisfied by a nearby node with lower latency. With the increasing number of users, the proposed caching strategy was able to utilize more cache resources in the network for data distribution, decreasing the overall latency with and without D2D communication.

% allowing the neighbours to fetch then with lower hops, whereas the other algorithms performed the same as the users increased. As such, the proposed distributed caching strategy will continue to perform better as the number of devices increases further. Moreover, even when no D2D communication was used, the proposed strategy was still able to provide substantial improvement over the default strategies.

%\subsection{Cache Hits}

\noindent \textbf{Cache hits:} Figures~\ref{fig:cache_hit} and~\ref{fig:cache_hit_d2d} present the number of Interests packets that were served by in-network caches without and with D2D communication respectively. As the number of F-UEs increases, the proposed caching strategy served more packets from in-network caches compared to other strategies. %The proposed cache distribution strategy distributes the in network caches optimally. 
Furthermore, the proposed caching strategy benefits from increasing the number of F-UEs in both cases (with and without D2D communication). 
% When D2D communication was available, each F-AP was able to use the available caches from its F-UEs to stasfy other F-UE's requests via the D2D technique. Moreover, the performance without D2D was not as good; however, it still outperformed other default strategies.

%\subsection{Fronthaul Channel Usage}

%When measuring the fronthaul channel usage, 
\noindent \textbf{Fronthaul channel usage:} We measured the number of packets that used the fronthaul channel to reach the central BBU pool. As shown in Figures~\ref{fig:fh} and~\ref{fig:fh_d2d}, the fronthaul channel usage decreased the most when D2D was enabled. As the number of F-UEs increases, more cache resources become available and the fronthaul usage decreases. %From this result, it can also be concluded that the proposed caching strategy will be able to reduce the backhaul channel usage as well.

% As such, it benefited from having more F-UEs than other strategies, and the fronthaul usage dropped as the number of users increased. On the other hand, the proposed cache distribution strategy still performed better when no D2D communication was used because of the optimal cache distribution among an F-AP and F-UE as seen in Figure~\ref{fig:fh_d2d}. From this result, it can also be concluded that the proposed caching strategy will be able to reduce the backhaul channel usage as well.

\begin{algorithm}
\small
  \caption{Caching algorithm.}\label{algo}
  \renewcommand{\algorithmicrequire}{\textbf{Input:}}
  \renewcommand{\algorithmicensure}{\textbf{Output:}}
  \begin{algorithmic}[1]
  \Require \textit{Interest packet} $i$, \textit{Data packet} $d$, refresh period $\tau$
  \Ensure \textit{Data packet $d$}
  \Statex \textit{Initialisation} : \textit{$t \leftarrow 0$, $\mathit{CS} \leftarrow \emptyset$, $\mathit{PIT} \leftarrow \emptyset$}
  \If{$i$ arrives}\label{algo1:interest_handling}
    \If{$d_i \in \mathit{CS}$}
        \Return $d_i$
    \ElsIf{$i \in \mathit{PIT}$} 
        \State Aggregate $i$.
    \Else
        \State Forward $i$ to the next node $n$.
        \State $\mathit{PIT} \leftarrow i$.
    \EndIf
\EndIf
\If{$d_i$ arrives}\label{algo1:data_arrive}
    \State Increase request rate $\lambda_d$.
    \If{$\mathit{CS}$ is not full}
        \State $\mathit{CS} \leftarrow d_{i}$.\label{algo1:check_cs}
    \Else
        \For{$\forall d \in \mathit{CS}$}
            \State Find $d_{min}$ with the minimal $\lambda$ and hops.\label{algo1:find_min}
        \EndFor
        \State Compare $d_{min}$ with $d_i$.
        \If{$d_{min}$ is smaller}
            \State Drop $d_{min}$ from $\mathit{CS}$.
            \State $\mathit{CS} \leftarrow d_{i}$.\label{algo1:replace_cs}
        \EndIf
    \EndIf
    \Return $d_i$
\EndIf

\If{$t = \tau$}
    \For{$\forall d \in \mathcal{D}$}
        \State Update request rate of $d$ using Equation~\eqref{eq:update_rate}.\label{algo1:update_rate}
    \EndFor
    \State $t \leftarrow 0$
\EndIf
\end{algorithmic}
\end{algorithm}

%% file: 6-conclusion.tex
\section{Conclusion}
\label{sec:con}

In this paper, we presented a design to incorporate NDN into the F-RAN architecture. Our main motivation was to reduce the burden on the fronthaul channel by allowing edge nodes to cache popular data. We proposed a caching strategy to achieve that and presented NDN-enabled designs for F-APs and F-UEs. Our simulation study showed that the proposed caching strategy can reduce the usage of the fronthaul channel, reduce the number of hops required for data retrieval, and increase cache hits as compared to baseline caching strategies. 

%The main motivation behind the design of the F-RAN architecture is reduce the burden on the fronthaul channel of the original C-RAN design by allowing edge nodes to cache popular data. By incorporating an ICN architecture like NDN, it is possible to further optimize this model. NDN has a native feature of in-network caching, which allows contents to automatically cache when a router forwards a Data packet. Modified design of F-AP and F-UE protocol stacks presented in this paper will help F-RAN to enable NDN support into the existing design. The proposed in-network cache distribution strategy allows F-RAN to optimally use the available cache resources. Simulation results demonstrate that the proposed caching strategy is capable of lowering the latency (average hops) and increasing the cache hit ratio. Furthermore, the proposed caching strategy is able reduce the burden on the fronthaul channel by optimally utilizing the available cache resources at the terminal layer of F-RAN. NDN is still at its infancy, and further researches are required to fully replace the well established IP network. The NDN enabled F-RAN architecture can be further redesigned to fully eliminated IP and use the NDN protocol solely. More efficient multi-path routing strategies can be proposed to efficiently transmit contents among the devices. Moreover, artificial Intelligence techniques such as federated learning can be incorporated to further optimize the NDN's caching strategy. All of these are regarded as future works.

%% file: apndx1.tex
\section{Proof of NP-Hardness of the Caching Problem}
\label{sec:apndx1}
\renewcommand{\theequation}{A.\arabic{equation}}
\setcounter{equation}{0}

To prove that the problem in \eqref{eq:obj_fn} is NP-hard, an F-RAN topology is considered that includes a central BBU pool $b \in \mathcal{A}$, two F-APs ($\{a1, a2\} \in \mathcal{A}$), and two F-UEs ($\{u1, u2\} \in \mathcal{U}$). $u1$ and $u2$ are served by $a1$ and $a2$ respectively. \textit{Interest packets} for content $c1$ and $c2$ are generated by $u1$ and $u2$ respectively.

Request rates at $a1$ and $a2$ are:
\begin{equation}
\begin{split}
    & \lambda_{c1}^{a1} = \lambda_{c1}^{u1}(1 - x_{c1}^{u1}), \quad
    \lambda_{c2}^{a2} = \lambda_{c2}^{u2}(1 - x_{c2}^{u2}).\\
\end{split}
\end{equation}

Request rates at $b$ are:
\begin{align}
    & \lambda_{c1}^{b} = \lambda_{c1}^{a1}(1 - x_{c1}^{a1}) =  \lambda_{c1}^{u1}(1 - x_{c1}^{u1})(1 - x_{c1}^{a1}),\\
    & \lambda_{c2}^{b} = \lambda_{c2}^{a2}(1 - x_{c2}^{a2}) =  \lambda_{c2}^{u2}(1 - x_{c2}^{u2})(1 - x_{c2}^{a2}).
\end{align}

We need to optimize the following:
\begin{align}\label{eq:np_prob}
    \notag
    \max \: & [\lambda_{c1}^{u1}x_{c1}^{b}(1 - x_{c1}^{u1})(1 - x_{c1}^{a1})\\\notag
    & + \lambda_{c2}^{u2}x_{c2}^{b}(1 - x_{c2}^{u2})(1 - x_{c2}^{a2})\\
    & + 2\lambda_{c1}^{u1}x_{c1}^{a1}(1 - x_{c1}^{u1}) + 2\lambda_{c2}^{u2}x_{c2}^{a2}(1 - x_{c2}^{u2})]\\\notag
    \textnormal{s. t. } & \sum x_{c1}^{b} + \sum x_{c2}^{b} \leq \mathit{CS}_b\\\notag
    & \sum x_{c1}^{a1} \leq \mathit{CS}_{a1}\\
    & \sum x_{c2}^{a2} \leq \mathit{CS}_{a2}.
\end{align}
%\begin{equation}
%\label{eq:np_prob}
%\begin{split}
%    \max \: & [\lambda_{c1}^{u1}x_{c1}^{b}(1 - x_{c1}^{u1})(1 - x_{c1}^{a1})\\
%    & + \lambda_{c2}^{u2}x_{c2}^{b}(1 - x_{c2}^{u2})(1 - x_{c2}^{a2})\\
%    & + 2\lambda_{c1}^{u1}x_{c1}^{a1}(1 - x_{c1}^{u1}) + 2\lambda_{c2}^{u2}x_{c2}^{a2}(1 - x_{c2}^{u2})]
%\end{split}
%\end{equation}
%\begin{equation}
%\begin{split}
%    \textnormal{s. t. } & \sum x_{c1}^{b} + \sum x_{c2}^{b} \leq \mathit{CS}_b\\
%    & \sum x_{c1}^{a1} \leq \mathit{CS}_{a1}\\
%    & \sum x_{c2}^{a2} \leq \mathit{CS}_{a2}.
%\end{split}
%\end{equation}
By rearranging~\eqref{eq:np_prob}, we have:
\begin{dmath}
\label{eq:np_prob2}
    \lambda_{c1}^{u1}[x_{c1}^b - x_{c1}^{a1}x_{c1}^{b} - x_{c1}^{u1}x_{c1}^{b} + x_{c1}^{u1}x_{c1}^{a1}x_{c1}^{b}]
    + \lambda_{c2}^{u2}[x_{c2}^b - x_{c2}^{a2}x_{c2}^{b} - x_{c2}^{u2}x_{c2}^{b} + x_{c2}^{u2}x_{c2}^{a2}x_{c2}^{b}]
    + 2\lambda_{c1}^{u1}[x_{c1}^{a1} - x_{c1}^{a1}x_{c1}^{u1}]
    + 2\lambda_{c2}^{u2}[x_{c2}^{a2} - x_{c2}^{a2}x_{c2}^{u2}],
\end{dmath}
where $x$ is a binary decision variable, which means some multiplications can be be replaced with the following variables given they satisfy the constrains:
\begin{equation}
\label{eq:z1}
z_{c1} =
\begin{cases}
x_{c1}^{a1}x_{c1}^{b} &\quad\quad \textit{s.t.} \, z_{c1} \leq x_{c1}^{b}\\
x_{c1}^{u1}x_{c1}^{b} &\quad\quad \textit{s.t.} \, z_{c1} \leq x_{c1}^{b}\\
x_{c1}^{a1}x_{c1}^{u1} &\quad\quad \textit{s.t.} \, z_{c1} \leq x_{c1}^{a1}\\
x_{c1}^{u1}x_{c1}^{a1}x_{c1}^{b} &\quad\quad \textit{s.t.} \, z_{c1} \leq x_{c1}^{b},
\end{cases}
\end{equation}
\begin{equation}
\label{eq:z2}
z_{c2} =
\begin{cases}
x_{c2}^{a2}x_{c2}^{b} &\quad\quad \textit{s.t.} \, z_{c2} \leq x_{c2}^{b}\\
x_{c2}^{u2}x_{c2}^{b} &\quad\quad \textit{s.t.} \, z_{c2} \leq x_{c2}^{b}\\
x_{c2}^{a2}x_{c2}^{u2} &\quad\quad \textit{s.t.} \, z_{c2} \leq x_{c2}^{a2}\\
x_{c2}^{u2}x_{c2}^{a2}x_{c2}^{b} &\quad\quad \textit{s.t.} \, z_{c2} \leq x_{c2}^{b}.
\end{cases}
\end{equation}
With the replacement of variables from~\eqref{eq:z1} and~\eqref{eq:z2}, \eqref{eq:np_prob2} can be rewritten as a zero-one linear programming problem, which is NP-hard:
\begin{dmath}
\label{eq:np_prob3}
    \lambda_{c1}^{u1}x_{c1}^b - \lambda_{c1}^{u1}z_{c1} + \lambda_{c2}^{u2}x_{c2}^b - \lambda_{c2}^{u2}z_{c2} 
    + 2\lambda_{c1}^{u1}x_{c1}^{a1} - 2\lambda_{c1}^{u1}z_{c1}
    + 2\lambda_{c2}^{u2}x_{c2}^{a2} - 2\lambda_{c2}^{u2}z_{c2}.
\end{dmath}